# Phonon origin of low energy and high energy kinks in high-temperature cuprate superconductors

E.A.Mazur[1(a)]

[1] National research nuclear university "MEPhI", Kashirskoe shosse, 31, Moscow, Russia

**Abstract -** Eliashberg theory generalized for the account of the electron-hole nonequivalence and electron correlations in the vertex function is used. The phonon contribution to the nodal electron Green function in cuprates is viewed. At non- zero temperatures the singularities (kinks) in the frequency behavior of a real and imaginary part of an electron nodal Green function, and also in the nodal part of the density of the electron states modified by an electron-phonon interaction are studied. It is shown that near the optimal doping both the low-energy and high-energy nodal Green function kinks and also the abnormal broadening of a band in cuprates are reproduced with the electron-phonon interaction in the extended Eliashberg theory.

In the number of recent experimental works [1–8], fulfilled by ARPES method [9,10], singularities (kinks) of the real and imaginary components of the self-energy (SE) part of an electron Green function (GF) in cuprates have been discovered at the low energy frequencies $\omega \sim 70-75 meV$. The correct interpretation of the origin of such kinks in the frequency SE part behavior and in the density of states in cuprates will allow us to reveal the mechanisms responsible for unusual properties of such materials in a normal state. Such a kink is also seen in electron doped compounds [11], while electron resonance mode is at 10 meV [12]. In the subsequent works (see, for example, [13–17]) besides the low-frequency singularities explored earlier in [1–7], the high-energy SE part singularities answering to the frequencies $\omega$ of the order 300-400 meV below a Fermi level have been discovered. In [17], it is revealed that the band width of the high-temperature superconductivity materials is wider than it is predicted with the evaluations without taking into account of the electron-phonon (EP) interaction. In [18,19], the isotopic dependence of the position of such singularities is discovered. This allows us to consider these kinks as a display in an electronic spectrum of the frequency singularities of the spectral function of the electron-phonon interaction $\alpha^2 F(\omega)$. The low-energy SE part singularity of the nodal GF is interpreted [1] as corresponding to the definite frequency of a phonon spectrum. The explanation of the occurrence of such kinks at high frequency $\omega$ several times exceeding the limiting phonon frequencies is still missing. Back in [20] it was shown by the present author with collaborators that the account of the difference from a constant of a density of states at the Fermi surface results in the reconstruction of the SE part and of the density of states in the EP system. The aim of present work is the investigation of the question what part of the experimental results is reconstructed with the EP interaction and what contribution herewith remains for the electron-electron interaction. The present work explores as well the problem whether the occurrence of the above mentioned kinks in the Green function SE part and in the density of states is caused by the EP interaction or by some additional electron mechanisms. For this purpose the extended variant of the Migdal-Eliashberg theory [20] under temperature $T \neq 0$ in Nambu representation is used which includes zone width finiteness and the inconstancy of the density of electron states and additionally the effects of the electron-hole nonequivalence.

Taking into account all written above we shall consider the EP system without Van-Hove singularity in the electron spectrum with Hamiltonian which includes the electron component $\hat{H}_e$, the ion component $\hat{H}_i$ and the component responding to the electron-ion interaction in the harmonic approximation $\hat{H}_{e-i}$

$\hat{H} = \hat{H}_e + \hat{H}_i + \hat{H}_{e-i} - \mu\hat{N}$. $\hat{H}_{ei}$ is represented with the following expression

$$\hat{H}_{ei} = \sum_{n\kappa} \int d\mathbf{r} \psi^+(x) \psi(x) \nabla_\alpha V_{ei\kappa}(\mathbf{r} - \mathbf{R}^0_{n\kappa}) u^\alpha_{n\kappa}$$

. Here the following notations are introduced: $M_\kappa$ as the mass of $\kappa$ - type ion, $\mu$ as the chemical potential, $\hat{N}$ as the operator of electron's number in the system, $u^\beta_{n\kappa}$ as an $\alpha$ - projection of a deviation of $\kappa, n$ ion from the balance position, $\mathbf{R}^0_{n\kappa} = \mathbf{R}^0_n + \boldsymbol{\rho}_\kappa$ as the radius-vector of the equilibrium position of $\kappa$ - type ion in the crystal, $V_{ei\kappa}$ as the potential of the electron-ion interaction, $\mathbf{u}_{n\kappa} = \mathbf{R}_{n\kappa} - \mathbf{R}^0_{n\kappa}$. The electron GF $\hat{G}$ in matrix Nambu form shall be defined as $\hat{G}(x, x') = -\langle T\Psi(x)\Psi^+(x')\rangle$ where the usual electron creation and destruction operators are entered in the form of Nambu operators. Writing down the standard movement equations for the electron wave functions and averaging with Hamiltonian $\hat{H}$ we'll be able to obtain the equation for the electron GF. The matrix SE part corresponding to the EP interaction with the coulomb vertex function $\hat{\Gamma}$ and full account of electron-electron correlations looks like

$$\hat{\Sigma}(x,x') = i\int dx_1 \int d\mathbf{r}'' \frac{e^2}{|\mathbf{r}-\mathbf{r}''|} \varepsilon^{-1}(\mathbf{r}''\tau, x_1) \hat{\tau}_3 \hat{G}(x, x_2) \hat{\tau}_3 \hat{\Gamma}(x_2, x', x_1) +$$
$$+i\left\{\sum_{n,\kappa;n',\kappa'} \int dx_1 dx_2 \nabla_\alpha V_{ei\kappa}(\mathbf{r} - \mathbf{R}^0_{n\kappa}) \nabla_\beta V_{ei\kappa}(\mathbf{r}_1 - \mathbf{R}^0_{n'\kappa'}) D^{\alpha\beta}_{n\kappa n'\kappa'}(\tau - \tau_1)\right\} \hat{\tau}_3 \hat{G}(x, x_2) \hat{\tau}_3 \hat{\Gamma}(x_2, x', x_1).$$
(1)

The phonon Green's function shall be defined as $D^{\alpha\beta}_{n\kappa,n'\kappa'}(\tau) = -i\langle T_\tau(u^\alpha_{n\kappa}, u^\beta_{n'\kappa'})\rangle$, $\hat{\Gamma}$ is the vertex function being a matrix in $\hat{\tau}_i$ space. The $\hat{\Gamma}$ behavior is formed under the influence of the first term in (1) which incorporates all the effects of the electron-electron correlations. In particular $\hat{\Gamma}$ is supposed to have the well known d-character on impulse in cuprates. Hereinafter we shall not draw the first contribution $\hat{\Sigma}_{el-el}(x, x')$ from (1) having it in mind and considering via behavior of $\hat{\Gamma}$ and $\hat{\Sigma}_{el-el}(x,x')$ all the revealed earlier (see, for example,[21]) effects of the electron-electron correlations and electron-magnon interaction in cuprates. After performance of the analytical continuation $i\omega_p \rightarrow \omega + i\delta$ the SE part of an electron Green function of the metal shall be represented as

$$\hat{\Sigma}^{ph}(p,\omega) = -\int \frac{d^3 p'}{(2\pi)^3} \sum_j |\Gamma(p,p')g_j(p,p')|^2 \int_{-\infty}^{+\infty} \frac{dz}{2\pi} \int_{-\infty}^{+\infty} \frac{dz'}{2\pi} \frac{th\frac{z'}{2T} + cth\frac{z}{2T}}{\omega - z - z' + i\delta} \hat{\tau}_3 \operatorname{Im} g_R(p',z')\hat{\tau}_3 b_j(p-p',z). \quad (2)$$

In (1) and (2) the Pauli matrixes $\hat{\tau}_i$ are introduced, $x \equiv \{r,t\}$, the direction of a spin is depicted by up and down arrows, frequency dependence in $\Gamma$ is neglected. In (2) $g_R$ is a retarded electron Green function, $b_j(p-p',z)$ is a spectral density of a phonon Green function,

$$g_j(p,p') = -\sum_\kappa \frac{1}{\sqrt{2NM_\kappa \omega_j(q)}} \langle p | e_j(q,\kappa) \nabla V_{ei\kappa}(r) | p' \rangle e^{iq\rho_\kappa}$$

, where $\omega_j(q)$ is a j-branch phonon frequency and $e_j(q,\kappa)$ is a polarisation vector corresponding to the impulse $q = p - p'$. Integration on an impulse shall be represented as follows $\int \frac{d^3 p}{(2\pi)^3}... = \int d\xi \int_{S(\xi)} \frac{d^2 p}{v_p}...$, where $v_p$ is the electron velocity modulo at the energy surface $\xi$. We shall take into account that near the optimal doping $\delta$ coulomb vertex $\Gamma(p,p')$ is greatly reduced due to the forward scattering peak (FSP) in the processes of the correlated electron scattering [22] for the transverse impulse component $q_\perp > \delta \cdot \pi / a$ with $a$ as a translation constant of the cuprates' plane. So the matrix element $\Gamma g_j(p,p')$ for impulse $p$ having nodal (antinodal) direction shall be assumed to have the following form defined with the use of $\theta$ function $\Gamma(p,p')g_j(p,p') = \Gamma g_{jnod}(\xi,\xi') \cdot (1 - \theta(|\varphi_{nod} - \varphi| - \delta \cdot \pi / aq(\xi)))$, thus supposing EP matrix element $\Gamma g_j(p,p')$ to be not dependent on an angle $\varphi$ indicating a deviation of $q$ from the nodal (antinodal) direction in the quasi 2D Fermi volume. In a vicinity of the optimum doping FSP is still distinctly expressed [22]. As a result we obtain separated equations for the nodal and antinodal Green functions in the neighborhood of the optimal doping $\delta \approx .16$. In the case of the final width of an electron band we shall consider the presence of bare (not renormalized with the EP scattering) variable fraction of the density of states $N_{0nod}(\xi)$ defined with the following expression $\int_{S(\xi)} \Gamma(p_{nod},p') \frac{d^2 p'}{v_{\xi p'}} d\xi = \int_{S(\xi)} \Gamma N_{0nod}(\xi) d\xi$ at the energy of bare electrons $\xi$ marked from a Fermi level with the electron impulse $p$ oriented in the nodal direction. We shall introduce a notation for the electron-phonon interaction spectral function as follows

$$\alpha_{nod}^2(\xi',\xi,z)F(\xi',\xi,z) = \frac{1}{2\pi}\int_{S(\xi')}\frac{d^2\boldsymbol{p}'}{v_{\boldsymbol{p}'}}\sum_j |\Gamma(\boldsymbol{p}_{nod},\boldsymbol{p}')g_j(\boldsymbol{p}_{nod},\boldsymbol{p}')|^2 b_j(\boldsymbol{p}_{nod}-\boldsymbol{p}',z)\times$$

$$\times\left(\int_{S(\xi')}\Gamma(\boldsymbol{p}_{nod},\boldsymbol{p}')\frac{d^2\boldsymbol{p}'}{v_{\boldsymbol{p}'}}\right)^{-1}$$

(3)

The SE part of the Green function shall be expressed with the use of (3) as follows

$$\hat{\Sigma}_{nod}^{ph}(\xi,\omega) = -\frac{1}{2\pi}\int_{-\infty}^{+\infty}dz\int_{-\infty}^{+\infty}dz'\int_{-\mu}^{+\infty}d\xi'\alpha_{nod}^2(\xi',\xi,z)F(\xi',\xi,z)\Gamma N_{0nod}(\xi')\frac{th\frac{z'}{2T}+cth\frac{z}{2T}}{\omega-z-z'+i\delta}\hat{\tau}_3\,\text{Im}\,\hat{g}_{Rnod}(\xi',z')\hat{\tau}_3.$$

(4)

The analogues equation is to be written for the antinodal impulse $\boldsymbol{p}$ direction so the EP interaction is assumed to be anisotropic [23]. In the definition (3) the vertex function $\Gamma$ is included that guarantees the inclusion of all the electron-electron correlation effects in the electron-phonon coupling description and the d-behavior of the electron-phonon self energy part. For the SE part the following decomposition on Pauli matrixes is written down $\hat{\Sigma}_{\varphi}^{ph}(\xi,\omega) = [1-Z_{\varphi}(\xi,\omega)]\omega\hat{\tau}_0 + \chi_{\varphi}(\xi,\omega)\hat{\tau}_3 + \varphi_{\varphi}(\xi,\omega)\hat{\tau}_1$. Here $\chi_{\varphi}(\xi,\omega)$ is a chemical potential renormalization in the $\varphi-$ direction due to the EP interaction. Impulses of electrons are not supposed to lie on the Fermi surface. Unlike [24] the integration is fulfilled in the entire Fermi volume and not just on the Fermi surface. Let's neglect in (4) the dependence of $\alpha_{\varphi}^2(\xi',\xi,z)F(\xi',\xi,z)$ from $\xi,\xi'$ values. We shall replace $Z(\vec{p}',\omega)$ with the quantity $Z_{\varphi}(\xi,\omega)$ corresponding to the constant energy $\xi$ in the direction defined with the angle $\varphi$. As a first approximation we shall neglect modification of the chemical potential $\mu$ due to the EP interaction thus supposing $\chi_{\varphi}(\xi,\omega)$ to be equal zero. At non-zero temperature taking into account the smallness of $\alpha^2(z)F(z)$ at low frequencies it let's assume that $cth\frac{z}{2T}\to 1$. From (4) taking into account the standard expression for the retarded GF $\hat{g}_{Rnod}(\xi',z')$ and neglecting dependence of $\text{Re}\,\Sigma_{\varphi}(\xi,\omega)$ as well as of $\text{Im}\,\Sigma_{\varphi}(\xi,\omega)$ on $\xi$ we obtain for the SE real part $\text{Re}\,\Sigma_{\varphi}(\omega) = [1-\text{Re}\,Z_{\varphi}(\omega)]\omega$ the following expression

$$[1-\operatorname{Re} Z_{nod}(\omega)]\omega = -P\int_0^{+\infty} dz\,\alpha_{nod}^2(z)F(z)\int_0^{+\infty} dz'\{f(-z')\times$$
$$\times\left(-\frac{N_{nod}(-z')}{z'+z+\omega}+\frac{N_{nod}(z')}{z'+z-\omega}\right)+f(z')\left(-\frac{N_{nod}(-z')}{z'-z+\omega}+\frac{N_{nod}(z')}{z'-z-\omega}\right)\}. \qquad (5)$$

Here $f(z')$ is the Fermi distribution function, an integral on z is taken in the sense of a principal value. From (5) it is evident that the SE part renormalization is not restricted by frequencies $\omega$ of the order of the limiting phonon frequency $\omega_D$ and spreads to essential energy $\omega \gg \omega_D$ as the SE real part at $\lambda \geq 1$ decreases at high frequency just as an inverse function of the frequency $\omega$. Therefore, the EP interaction modifies Green function SE part on considerable depth from the Fermi surface in units of the Debye phonon frequencies, and not just in the neighborhood of a Fermi surface $\mu - \omega_D < \omega < \mu + \omega_D$. For the SE imaginary part $\operatorname{Im}\Sigma_\varphi(\omega) = -\operatorname{Im} Z_\varphi(\omega)\omega$ we obtain as a result of straight lines but lengthy calculations the expression as follows

$$\operatorname{Im} Z_{nod}(\omega)\omega = \pi\int_0^{+\infty} dz\,\alpha_{nod}^2(z)F(z)\times$$
$$\times\{[N_{nod}(\omega-z)+N_{nod}(\omega+z)]n_B(z)+N_{nod}(\omega-z)f(z-\omega)+N_{nod}(\omega+z)f(z+\omega)\}. \qquad (6)$$

In (6), $n_B(z)$ is the Bose distribution function. In (5) and (6) the renormalized with the EP interaction nodal part of the density of electron states $N_{nod}(z')$ is expressed through the "bare" partial nodal density of electron states $N_{0nod}(\xi)$

$$N_{nod}(z') = -\frac{1}{\pi}\int_{-\mu}^{\infty} d\xi'\Gamma N_{0nod}(\xi')\operatorname{Im} g_{Rnod}(\xi',z'). \qquad (7)$$

In the following text subscript "nod" shall be omitted or replaced for the brevity with the subscript "n". Strictly speaking such density of states as $N(z')$ is spotted within the following range $-\infty < z' < +\infty$ and is not the symmetrical (even) function of $z'$. We will approximate further the expression for the bare electron density of states $\Gamma N_0(\xi)$ with $\Gamma N_0(\xi) = N_0$ at $-\mu \leq \xi \leq 2W-\mu$ where $W$ is the halfwidth of the initial bare band. In the remaining field of the variable $\xi$ the bare density of states $\Gamma N_0(\xi)$ shall be assumed to be equal to zero. We will consider for simplicity that the hole or electron type doping shifts the chemical potential $\mu$ from the half filling value in a linear mode as a function of a doping degree $\delta$ so that $\mu = W(1+\delta)$. Such a behavior of the chemical potential with doping is confirmed on experiment [25] and allows us to consider a doping degree $\delta$ of electron or of the hole- type in parts from the halfwidth of a band. We shall take into account the fact that in the neighborhood of the nodal direction an anomalous Green function can be assumed to equal zero. In our calculations we shall model experimentally observed spectral function $\alpha^2 F(\omega)\cdot N_0$ with the piece-wise function "adjusted" to the spectral function $\alpha^2 F(z)$ for one of the cuprates' representative Bi2212 ($Bi_2Sr_2CaCu_2O_{8+\delta}$) [26–28]. Spectral function of the EP interaction brought on Fig.1 is peculiar to the cuprates [26–30]. Some calculations of the potential of the EP interaction [31,32] predict much smaller values for the EP interaction constant than results from Fig.1. In a recent rather precisional experimental work [33] it is shown that calculations [31,32] disagree with the experimental results and shall be revised. In [34] with the inversion method the EP interaction constant values are received being in accordance with [26–28] in the range of the phonon frequencies. The other calculations [22] (see as well [35] and ref. therein) give significantly higher values of the EP interaction constant than received in [31,32]. The conclusion about unfairness of the approach [31] is made as well in [36]. In [36] it was shown that calculations employed in [31] fail to reproduce huge influence of EP coupling on important phonons observed in experiments. It was stressed in [36] that such an increase of EP coupling in comparison with [31,32] results of enhanced electronic correlations in cuprates not included in DFT. Let's assume $\alpha^2 F(z)$ to be a function of the dimensionless frequency $z = \omega/\omega_D$ as shown below $\alpha^2 F(\omega)\cdot N_0 =$ 0 at 0 <z <18/75; 0.5 at 18/75 <z <43/75; 0.45 at 43/75 <z <55/75, .65 at 55/75<z<1; 4.43-3.8z at 1 <z <1+1/6; 0 at z> 1+1/6. Such adjusted expression corresponds to the above mentioned featured view of $\alpha^2 F(z)$ (Fig.1). This allows us to calculate analytically one of the integrals in (6) facilitating calculations with the single remaining integral. In the case of the finite width of the bare electron band when $\Gamma N_0(\xi)$ differs from zero in a finite interval of the variable $\xi$ it immediately becomes clear from (6) that SE real part $\operatorname{Re} Z(z')$ and consequently SE imaginary part $\operatorname{Im} Z(z')$ due to the dispersing relations as well as the density of states $N(z')$ (as a result of (5) - (7)) all differ from zero in a wider interval of frequencies than the bare density of states $N_0(\xi)$. Thus the EP interaction does not narrow down an electron band and on the contrary increases the width of a band with simultaneous redistribution of the density of states in this band. This very effect is observed in the experiment in cuprates [17] and proves to be true as shown in the present work involving math calculations. The difference of the chemical potential position from the symmetrical in an electron band of the doped material in a combination with the deep reconstruction of the Fermi volume results in the occurrence of the frequency asymmetric behavior of the functions $\omega \operatorname{Im} Z(\omega)$ and $\omega \operatorname{Re} Z(\omega)$ in the ranges of the negative and positive frequencies (an electron-hole asymmetry). In the present work we shall restrict our consideration to iterative approach to the solution of combined equations (5), (6) and (7). Results of calculations for the SE part of the nodal GF are presented (Fig.2) as graphs of functions of the dimensionless parameters $\lambda$, $W/\omega_D$ and $T/\omega_D$. Fig.2 shows singularities of $\operatorname{Re}\Sigma(\omega)$, $\operatorname{Im}\Sigma(\omega)$ and also of $N(\omega)$ resulting from extended in [20] and in the Eliashberg theory. The use of such generalization [20] of the Eliashberg theory within the self consistent scheme and with FSP has allowed us to study the effects of the finiteness of an electron band width, the electron-hole nonequivalence and the renormalization of the density of electron states $N(\omega)$ in the different impulse directions.

On the Fig.2 the results of the calculations for $W/\omega_D = 5$ [10], T =. 15 for 2212 are presented. The revealed in the calculations low-energy and the high-energy kinks (HEK) in the SE part and also in the density of states are marked with the arrows. As it becomes evident from (7) as well as from the bare density of the states $\Gamma N_{0nod}(\xi)$ definition the HEK in the nodal fraction of the density of states may be interpreted as a singularity of the nodal electron velocity that is seen in experiments [13 – 17]. On the Fig. 2 (b) and (c) an extremely small difference from zero of $\text{Im}\Sigma(\omega)$ and $N(\omega)$ at the frequency exceeding the renormalized zone width is not shown. At the frequencies smaller than 1 the SE real part behavior may be characterized as a "shoulder" observed in experiments [1,2,14]. It becomes clear that the density of states and also the electron spectrum show reconstruction at frequencies up to several dimensionless units. The results of the calculations of the $\text{Re}\Sigma(\omega)$ in the case of the $\delta$ =0.16 hole-type optimal doping at $T = 130K$ with the use of $\alpha^2 F(z)$ which models the experimentally observed $\alpha^2 F(z)$ [26 – 28] for 2212 are built in top of the experimental plots for the SE real part $\text{Re}\Sigma(\omega)$ from the work [34] for 2212 (FIG.3). In the frequency range where $\text{Re}\Sigma(\omega)$ is negative ARPES experiments are in present practically absent (for the exception of [37] where not quite definite results are presented). FIG.3 clearly shows that the calculated values of the $\text{Re}\Sigma(\omega)$ in the frequency range important for the superconducting pairing $\omega < 75meV$ coincide with the SE real part $\text{Re}\Sigma(\omega)$ obtained in the experiment [34,44]. The EP interaction constant is evaluated as $\lambda \sim 1.21$ as the slope of $\text{Re}\Sigma(\omega) \sim -\lambda\omega$ on FIG.3 at low frequency that agrees with the estimates [38,39] and is found to be in full consent with the results presented in reviews [40,41] .Thereby, the contribution from the EP interaction to the nodal GF self-energy part is calculated in the present work with the use of the self-consistent scheme of the calculation at non zero temperature conditioned with the electron-electron correlation in the form of the FSP with account of the zone width finiteness, chemical potential modification with doping and the electron-hole nonequivalence. The calculation does not contain any fitting parameters since the spectral function of the EP interaction starts from the tunnel measurements. The fact that electrons are considered in all Fermi volume rather than only "lying" on the Fermi surface allows us to describe the reconstruction of the Fermi volume due to the electron spectrum renormalization with cascade of phonons "feeling" zone bottom in agreement with experiments [42,43]. Summing up all written above we arrive to the following conclusions: 1. Calculated in the present work electron nodal Green function SE part values for the typical representative of cuprates Bi2212 with the use of the adjusted to the experimental curve model EP spectral function coincide with the experimental data [34,44] in the frequency range not exceeding typical energies of the phonon spectrum. The account of the electron zone width finiteness brings the inversion of $\text{Re}\Sigma_{nod}(\omega)$ to zero at energy $\omega_0 \approx 400 - 420meV$ complying with values observed in experiments [34,44]. For reproducing such an energy $\omega_0$ there is no need to assume a presence of a high energy "tail" in the electron-boson interaction spectral functions [34,45 – 48]. The present consideration does not reduce the GF description to the interaction of an electron with phonons only since calculations called in the present work use FSP being a many body electron effect. Correctly describing behavior of $\text{Re}\Sigma(\omega)$ at frequency not exceeding frequency of the phonon spectrum, reproducing the energy $\omega_0 \approx 420 meV$ of the inversion of $\text{Re}\Sigma(\omega)$ to zero as well as the positions of the LEK and HEK, calculated phonon contribution $\text{Re}\Sigma(\omega)$ significantly yields in magnitude to the experimental values [34, 44] in the range of frequency exceeding typical phonon frequency. It's necessary to refer to the possible reasons of the mismatch with the experiment of the calculated curve in the high energy range the inexactness caused with the approximate description of the vertex function $\Gamma^{sf}$ answering EP interaction complicated with electron correlation in the form of the FSP. Less probably that the uncertainty in the electron bare dispersion or a neglect of the inconstancy of $N_{d0}(\varepsilon)$ in calculations, in particular, neglect of the Van-Hove singularities answers for this mismatch. This mismatch cannot be explained with the neglect of the high energy tail in the spectral function of the electron-boson interaction answering to the electron interaction with the spin fluctuations (SF), since such a situation would disagree with experiment [49]. In [49] the strong evidence that the SF pairing mechanism is not the dominating one in HTSC cuprates is represented with the revealed in this work pronounced rearrangement and suppression of $\text{Im}\chi^{(odd)}(Q,\omega)$ (in the normal state of cuprate) by doping toward the optimal doping with simultaneous negligible change in $T_c$. 2. $\text{Re}\Sigma(\omega)$ and $\text{Im}\Sigma(\omega)$ have values of the one order as it is seen from Fig.2 in full accordance with the experimental results [44, 50]. This fact makes it impossible to ignore one of the values $\text{Re}\Sigma(\omega)$ or $\text{Im}\Sigma(\omega)$ in the process of the inverse problem resolving as it is usually done [46 – 48, 51]. So the appearance of the EF with high energy tails received within the framework of such approach [34, 46 – 48] has not a due motivation. 3. High-energy kink (HEK) in the nodal part of the electron density $N_n(\omega)$, real $\text{Re}\Sigma_n(\omega)$ and imaginary $\text{Im}_n\Sigma(\omega)$ parts of the electron self energy is discovered in the present work (Fig.2). Nodal part of the electron density $N_n(\omega)$ is shown to be related with the electron velocity (electron dispersion singularities) in the nodal part of the Fermi volume. As a result it's shown in the present paper that HEK discovered in [13 – 17] is also bound by the origin to the EP interaction at $\lambda \sim 1$ and to the singularities in the bare electron spectrum. The last conclusion does not disagree with the fact that the energy of such kink $\omega \sim 300 - 400 meV$ essentially exceeds the characteristic phonon energy $\omega \sim 70 - 80 meV$. HEK reveals the significance of the multiphonon processes and is explained by the weak reduction of the SE part at such processes that gives the EP system an opportunity to "feel" singularities or zone bottom in an electron spectrum at the large energies measured from the Fermi surface in accordance with the experimental works [42, 43]. Thus it is shown that both the low-frequency and the high-frequency SE part singularities (kinks) in cuprates originate from the EP interaction and to the peculiarities in the bare electron spectrum. 5. The effect of the electron band broadening with the EP interaction is established and confirmed by numerical calculations at the EP interaction constant $\lambda \geq 1$. The proof of the effect of the broadening of the electron band width with the EP interaction allows us to explain the effect observed experimentally in cuprates [17]. The mentioned effect didn't find any explanation before.

The author thanks Yu.Kagan, A.S. Alexandrov, V.F.Elesin, E.A.Manjikin, A.I.Agafonov, N.A.Kudrjashov, N.B. Narozhnjij, D.N.Voskresenskij, H.S. Ruiz for the discussions of the present work.

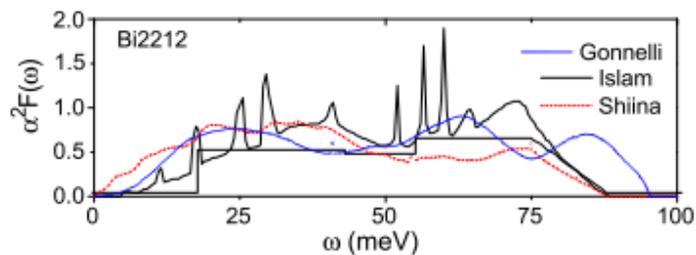

Fig.1 Experimental electron−phonon spectral function $\alpha^2F(\omega)$ [24,25,26] in comparison with the adjusted $\alpha^2F(\omega)$ (thin solid line)

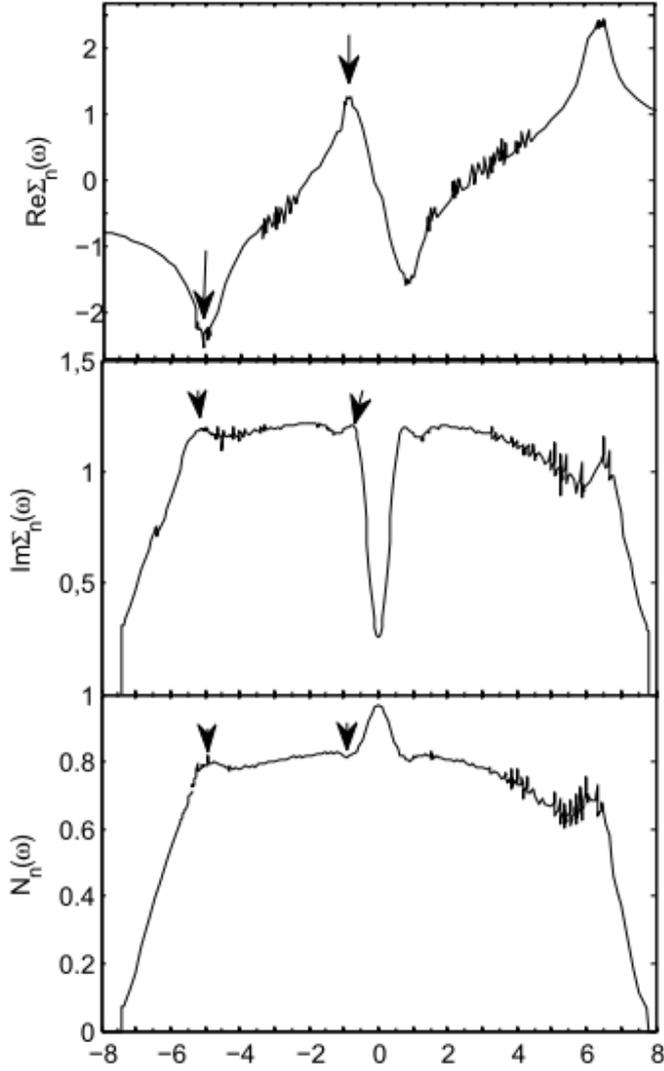

FIG.2. Real part (a) of the self energy $Re\Sigma_n(\omega)$, (b) imaginary part of the self energy $Im\Sigma_n(\omega)$, (c) electron density of states $N_n(\varepsilon)$, renormalized with the EP interaction vs. energy $\omega$ below and up the Fermi level. $\omega$, $Re\Sigma_n(\omega)$, $Im\Sigma_n(\omega)$, temperature T, chemical potential $\mu$, zone halfwidth W are represented in dimensionless units as described i the text, $N_n(\varepsilon)$ is renormalized with the initial density $N_0$, T=.15. Low energy and high energy kinks are marked with the arrows. Calculations performed for the optimally doped Bi2212, $\delta$=.16 ( hole doping).

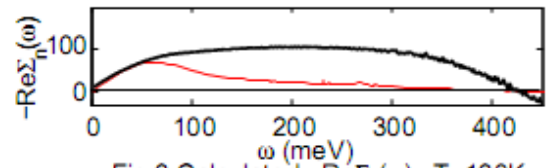

Fig.3. Calculated $-Re\Sigma_n(\omega)$, T=130K, (thin black solid line) in comparison with the experimental $-Re\Sigma_n(\omega)$ for the optimally doped Bi2212 [32].